\documentstyle[prb,aps,twocolumn]{revtex}
\begin{document}

\twocolumn[\hsize\textwidth\columnwidth\hsize\csname 
@twocolumnfalse\endcsname

\title{Staggered pairing phenomenology for UPd$_2$Al$_3$ and
       UNi$_2$Al$_3$}
\author{V. Martisovits}
\address{Department of Physics, Ohio State University, Columbus,
         Ohio 43210}
\author{D. L. Cox}
\address{Department of Physics, Ohio State University, Columbus,
         Ohio 43210\\ and Department of Physics, University of 
         California, Davis, California 95616\cite{addr}} 
\maketitle

\widetext

\begin{abstract}
We apply the staggered-pairing Ginzburg-Landau phenomenology to describe 
superconductivity in UPd$_2$Al$_3$ and UNi$_2$Al$_3$. The phenomenology was 
applied successfully to UPt$_3$ so it explains why these materials have 
qualitatively different superconducting phase diagrams although they have 
the same point-group symmetry. UPd$_2$Al$_3$ and UNi$_2$Al$_3$ have a
two-component superconducting order parameter transforming as an $H$-point 
irreducible representation of the space group. Staggered superconductivity 
can induce charge-density waves characterized by new Bragg peaks suggesting 
experimental tests of the phenomenology.
\end{abstract}


  ]

\narrowtext

Heavy-fermion superconductors\cite{grewe,hess}
exhibit many signs of unconventional superconductivity thus playing an 
important role in advancing our knowledge of this important subject. 
One of the clearest 
signs of this unconventional pairing 
is a complex phase diagram with multiple superconducting phases. 
This is clearly the case for UPt$_3$ which has three distinct 
superconducting phases.\cite{bruls,adenwalla,boukhny} On the other hand,
UPd$_2$Al$_3$ and UNi$_2$Al$_3$ have simple phase diagrams with only one 
superconducting phase.\cite{geibel} The 
crystal structures of UPt$_3$, UPd$_2$Al$_3$, and UNi$_2$Al$_3$ are 
characterized by the {\em same\/} point group ($D_{6h}$) and the
{\em same\/} Bravais lattice (simple hexagonal) with {\em different\/} numbers 
of uranium atoms in the basis (two for UPt$_3$, one for UPd$_2$Al$_3$ and   
UNi$_2$Al$_3$).

So far no comprehensive microscopic theory exists for these materials. Many 
phenomenological approaches\cite{hess2,ohmi,chen,sauls} 
have tried to explain exotic superconductivity 
in UPt$_3$. Most of them use a multidimensional superconducting order 
parameter transforming as a representation of the point group $D_{6h}$ and 
a Ginzburg-Landau free energy invariant under this group. If one tries to 
apply this approach to describe superconductivity in UPd$_2$Al$_3$ and      
UNi$_2$Al$_3$, it is necessary to choose an order parameter transforming as 
a different representation of the group $D_{6h}$. No single phenomenology 
is able to predict satisfactorily the relevant representation for each 
material.

In this paper, we apply the staggered-superconductivity Ginzburg-Landau 
theory\cite{heid,heid2}
to describe superconductivity in UPd$_2$Al$_3$ and UNi$_2$Al$_3$. 
This theory has been successful in explaining the $H$-$P$-$T$ 
superconducting phase diagram of UPt$_3$\cite{bruls,adenwalla,boukhny}
including the tetracritical point 
for all directions of the magnetic field. The distinctive feature of this 
theory is a superconducting order parameter transforming as an 
irreducible representation of the space group 
characterizing the crystal structure which resides {\it away} from the
center of the Brillouin zone. Namely, we have finite center-of-mass or 
{\it staggered} pairing.  For UPd$_2$Al$_3$ and 
UNi$_2$Al$_3$, the star in the reciprocal space carrying the representation 
is determined by the negative pair hopping model. We introduce a  
two-component $H$-point superconducting order parameter and write a free 
energy including gradient terms and coupling to a symmetry-breaking strain. 
Minimizing the free energy in the absence of the magnetic field, we 
find stability domains in the parameter space, and identify 
sequences of possible 
phases for each domain. Application of an external or spontaneous 
symmetry-breaking strain does not split the 
superconducting transition. The calculated upper critical field as a 
function of the temperature shows no kink. We introduce 
charge-density-wave and magnetic order parameters and investigate their 
possible coupling to the superconducting order parameter. 
Charge-density-wave order can be induced by staggered superconductivity 
suggesting a possible experimental test of the phenomenmology. 
On the other hand, magnetic order 
cannot be induced.

The motivation for the staggered-superconductivity theory is based upon 
the theory of 
odd-in-frequency pairing characterized by a gap function which is an odd 
function of the frequency.\cite{balatsky,balatsky2}   This motivation is 
two-fold: (i) there are strong reasons to believe that an
odd-in-frequency superconducting state cannot be stable for pairs with
zero center-of-mass momentum,\cite{coleman,heid3} and (ii) the two-channel
Kondo lattice, which is one candidate theory for heavy fermion
compounds, has recently been shown to produce odd-frequency
superconductivity in infinite spatial dimensions\cite{jarpangcox} which 
is likely to be staggered for finite dimensionality.  (For a recent
review of two-channel Kondo lattice models and related physics, see
\cite{coxzow}.)

In consequence, we develop a phenomenology for a
non-uniform superconducting state 
described by a staggered superconducting order parameter and a negative 
pair hopping energy between uranium sites.\cite{coleman} 
We then write a hopping model 
for Cooper pairs assuming in-plane and out-of-plane nearest 
neighbors. The energy band for this model has a two-fold degenerate 
minimum for the crystal momenta located at the opposite corners 
$H_1 , H_2$ of the Brillouin zone shown in Fig.\ \ref{BZ}.  
(For UPt$_3$, the different number of 
uranium atoms in the basis moves the minimum to the $M$ 
points.\cite{heid}) Cooper 
pairs will condense in states with these momenta so in this picture there 
are pairs with finite center-of-mass momenta. The Bloch functions
$\psi_1({\bf r}) , \psi_2({\bf r})$ of these states are uniquely determined 
by specifying the Wannier function. We assume that it transforms as a
real one-dimensional representation of the uranium-site point group
$D_{6h}$.  For example, it will transform as 
$A_2$ if we use a two-channel quadrupolar Kondo model to describe
UPd$_2$Al$_3$.\cite{coxzow} However, all the following results do not depend 
on this specific choice.

We introduce a two-component superconducting order parameter
$\eta_1 , \eta_2$ using the gap function $\Delta ({\bf r})$\cite{oddgap} 
where ${\bf r}$ is the pair center-of-mass coordinate:
\begin{equation}
 \Delta({\bf r}) = (\eta_1-i\eta_2) \psi_1({\bf r}) +
                   (\eta_1+i\eta_2) \psi_2({\bf r}) .
\end{equation}
The choice of the coefficients in front of the Bloch functions gives the 
usual transformation properties under the time reversal transformation
$\eta_1 \rightarrow \eta_1^*$, $\eta_2 \rightarrow \eta_2^*$. With respect 
to the space group $D_{6h}^1$, the order parameter transforms as a 
two-dimensional $H$-point irreducible representation. 

The free-energy density $F$ is required to be invariant under the elements 
of the space group $D_{6h}^1$, the $U(1)$ gauge transformations, and 
time reversal. We include coupling to the hexagonal-symmetry-breaking 
strains $\epsilon_1 \sim x^2-y^2$, $\epsilon_2 \sim 2xy$ which transform as 
the center-of-Brillouin-zone representation $E_{2g}$. The bulk part $F_0$ 
and the gradient part $F_g$ are given by
\begin{mathletters}
\begin{equation}
 F = F_0 + F_g , 
\end{equation}
\begin{eqnarray}
 F_0 & = & \alpha(T) \left(|\eta_1|^2+|\eta_2|^2\right) + 
          \beta_1   \left(|\eta_1|^2+|\eta_2|^2\right)^2 \nonumber \\
  && \mbox{} + \beta_2 \left(|\eta_1|^4+|\eta_2|^4+\eta_1^2 \eta_2^{*2}+
                        \eta_1^{*2} \eta_2^2 \right) ,  
\end{eqnarray}
\begin{eqnarray}
 F_g & = & \kappa \sum_n \left(|p_x\eta_n|^2 
          + |p_y\eta_n|^2 \right) \nonumber \\
  &&   \mbox{}   +  \kappa_\epsilon \epsilon_1 
         \sum_n \left(|p_x\eta_n|^2 - |p_y\eta_n|^2 \right) \nonumber \\
  & & \mbox{} + \kappa_\epsilon \epsilon_2
   \sum_n \left[(p_x\eta_n)(p_y\eta_n)^*+(p_x\eta_n)^*(p_y\eta_n)\right]
   \nonumber \\
  & & \mbox{} + \kappa_z \sum_n |p_z\eta_n|^2 .
\end{eqnarray}
\end{mathletters}
Here $\alpha(T)=a_0(T-T_c)$ with $T_c = T_{c0} - \left(\epsilon_1^2 +
\epsilon_2^2 \right)/a_0$; $\beta_1$, $\beta_2$, $\kappa$, 
$\kappa_\epsilon$, and $\kappa_z$ are phenomenological Ginzburg-Landau 
parameters; $p_\alpha=-i\partial_\alpha-2e A_\alpha /\hbar c$
($e<0$). $|\eta_1|^2$ and $|\eta_2|^2$ enter the quadratic term in the bulk 
part $F_0$ with the same coefficient $\alpha(T)$ which is valid up to any 
order in the strains $\epsilon_1,\epsilon_2$ (the above expression 
demonstrating this explicitly up to the second order). This implies that 
the presence of any 
finite symmetry-breaking strains do not split the superconducting 
transition which is in agreement with experiment in that only one 
superconducting transition is found (despite the presence of magnetic
order at elevated temperatures which, at least for UPd$_2$Al$_3$ is not
of the small moment variety found in UPt$_3$).\cite{geibel}

Minimizing the free energy, we can investigate possible superconducting 
phases for $T<T_c$ in the absence of the magnetic field. There are two 
regions of stability in the parameter space $\beta_1,\beta_2$ in which the 
free energy has a stable lower bound. 
For the region I defined as $\beta_1>0$,
$\beta_2>0$ there is only one phase $(\eta_1,\eta_2)=(i,1)$ breaking the 
symmetry under the time reversal (as favored by the last term in the bulk 
part $F_0$). The region II defined as $\beta_1+\beta_2>0$, $\beta_2<0$ is 
characterized by the phase $(\eta_1,\eta_2)=(0,1)$ having only one 
component of the order parameter condensed.\cite{sixth}

Let us now discuss the upper critical field $H_{c2}$. Minimizing the
free-energy functional with respect to the components of the 
superconducting order parameter and keeping only terms linear in them, we 
get two {\em identical\/} decoupled linearized Ginzburg-Landau equations. 
This means that the field $H_{c2}$ does not have any kink as a function of 
the temperature $T$ as observed experimentally
(for UPt$_3$ there are three {\em different\/} decoupled 
equations, and the field $H_{c2}$ is given by the largest of the three 
values generated by each equation so we get a kink corresponding to the 
tetracritical point\cite{heid}). The dependence of $H_{c2}$ on 
the direction of the 
magnetic field has the same 
form as the result for a conventional free energy with 
effective-mass anisotropy:
\begin{mathletters}
\begin{equation}
 H_{c2}(T)  = - \frac{\hbar c}{2|e|} \; \frac{\alpha(T)}{\sqrt{R}} ,
\end{equation}
\begin{equation}
 R  =  \left(\kappa^2-\kappa_\epsilon^2|\epsilon|^2\right)
        \cos^2\theta+\kappa_z \left[\kappa-\kappa_\epsilon
        \text{Re}\left(\epsilon e^{i2\varphi}\right)\right]
        \sin^2\theta ,
\end{equation}
\end{mathletters}
with $\epsilon=\epsilon_1-i\epsilon_2$.

Finally, let us investigate the possibility of charge-density waves and 
magnetic order induced by staggered superconductivity. Starting from the 
staggered gap function (1), we can introduce order parameters for them in a 
very general way.\cite{toledano}
The gap function $\Delta ({\bf r})$ does not have the 
periodicity of the original simple hexagonal Bravais lattice; it has the 
periodicity of a new larger simple hexagonal Bravais lattice of 
Fig.\ \ref{plane} with  
$a^\prime = \sqrt{3} a$, $c^\prime = 2c$. Thus the most general charge and 
magnetic structures are described by this new lattice and a basis 
consisting of six uranium atoms. Assigning different charges and magnetic 
moments for each element of the basis, we define six-component order 
parameters. The six components can be replaced by three 
interplanar-bonding combinations and three interplanar-antibonding ones. To 
simplify the following discussion, we can ignore the bonding 
combinations because they do not couple to the superconducting order 
parameter. Then the basis atoms in a plane and the basis atoms in a 
neighboring plane have the same charges $\rho_1,\rho_2,\rho_3$ and the 
same magnetic moments 
$  \bbox{\mu}_1 , \bbox{\mu}_2 ,  \bbox{\mu}_3  $.

The components $\rho_1,\rho_2,\rho_3$ of the charge order parameter 
transform under the space group $D^1_{6h}$ as a three-dimensional reducible 
representation composed of the totally symmetric irreducible representation 
and a two-dimensional $K$-point irreducible representation 
(Fig.\ \ref{BZ}). The totally 
symmetric combination $\rho_1+\rho_2+\rho_3$ represents the total charge in 
the large primitive cell which can be put equal to zero in the context of 
charge-density waves. Then the relevant part of the free-energy density is
\begin{eqnarray}
 F_{\text{cdw}} & =& \alpha_\rho \left(\rho_1^2+\rho_2^2+\rho_3^2-
           \rho_1\rho_2 -
           \rho_1\rho_3 - \rho_2\rho_3 \right)  \nonumber \\
 && \mbox{} +g \left(|\eta_2|^2-|\eta_1|^2\right)(2\rho_1-\rho_2-\rho_3)
        \nonumber \\
 && \mbox{}   + \sqrt{3}g\left(\eta_1\eta_2^*+\eta_1^*\eta_2\right)
         (\rho_2 - \rho_3)
\end{eqnarray}
where $\alpha_\rho$ and $g$ are phenomenological Ginzburg-Landau 
parameters. Minimizing this free energy, we find possible 
charge-density-wave structures for all the superconducting phases 
identified above. There is no induced charge-density wave for the phase
$(\eta_1,\eta_2)=(i,1)$ at the region I\@. For the phase 
$(\eta_1,\eta_2)=(0,1)$ at the region II, staggered superconductivity 
induces a charge-density wave described by the order parameter
$(\rho_1,\rho_2,\rho_3) = (2,-1,-1)$. The structure is characterized by new 
Bragg peaks $\left(\frac{1}{3} n_1, \frac{1}{3} n_1 +n_2, n_3\right)$ where
$n_1$, $n_2$, and $n_3$ are any integers. This important experimental 
prediction can be verified using neutron or x-ray scattering.
 
The components $  \bbox{\mu}_1 , \bbox{\mu}_2 ,  \bbox{\mu}_3  $
of the magnetic 
order parameter transform under the space group as the product of the
axial-vector representation and the representation for the charge order 
parameter discussed above. The product is a nine-dimensional reducible 
representation composed of two center-of-Brillouin-zone irreducible 
representations and two $K$-point irreducible representations. In the free 
energy there are no invariants coupling the magnetic order parameter 
linearly to the square of the superconducting order parameter. Staggered 
superconductivity does not induce magnetic order (either in-plane or
perpendicular-to-plane).

In summary, the Ginzburg-Landau theory of staggered superconductivity was 
applied to UPd$_2$Al$_3$ and UNi$_2$Al$_3$. Motivated by
odd-in-frequency pairing, we found the minima of the negative hopping model 
for Cooper pairs at the corners $H_1 , H_2$ of the Brillouin zone. We 
introduced a two-component superconducting order parameter
$\eta_1 , \eta_2$ transforming as an $H$-point irreducible representation 
of the space group. We constructed the free energy including gradient terms 
and coupling to symmetry-breaking strains. The superconducting transition 
is not split by any finite value of the strains. In the parameter space 
there are two zero-magnetic-field regions of stability characterized by the 
superconducting phases $(\eta_1,\eta_2)=(i,1)$ and $(0,1)$, respectively. 
The upper critical field as a function of the temperature has no kink. We 
defined charge and magnetic order parameters and investigated whether the 
superconducting order parameter can induce them. For the phase
$(\eta_1,\eta_2)=(0,1)$ there is an induced charge-density wave 
characterized by new Bragg peaks 
$\left(\frac{1}{3} n_1, \frac{1}{3} n_1 +n_2, n_3\right)$ which could be 
seen by neutron or x-ray scattering.
   
We acknowledge useful discussions with F. B. Anders, R. Heid, T.-L. Ho,
and H. R. Krishnamurthy. This research was supported by the National 
Science Foundation under Grant No.\ DMR-9420920. The part of this research 
done at the Institute for Theoretical Physics, University of California, 
Santa Barbara, California was supported by the National Science Foundation 
under Grant No.\ PHY-9407194.

\begin{figure}
\caption{Brillouin zone for the simple hexagonal Bravais lattice and 
elements of stars for representations characterizing different order 
parameters. The stars $\{ M_1, M_2, M_3 \}$ and $\{ H_1, H_2 \}$ correspond 
to the superconducting order parameters for UPt$_3$ and UPd$_2$Al$_3$, 
respectively. The stars $\{ \Gamma \}$ and $\{ K_1, K_2 \}$ describe the 
charge and the magnetic order parameters for UPd$_2$Al$_3$.}
\label{BZ}
\end{figure}

\begin{figure}
\caption{Original and cell-tripled hexagonal Bravais lattices (with only one 
plane shown) used to define the charge and the magnetic order parameters. 
The original lattice consists of all the points drawn. The new lattice 
consists of the points on the dotted lines and describes the most general 
charge and magnetic structures. Points characterized by different symbols 
represent uranium atoms with different charges and magnetic moments.}
\label{plane}
\end{figure}


\begin{references}
\bibitem{oddgap}If our staggered order parameter arises from
  odd-frequency pairs, it must be understood to correspond to the
  expectation value of the first derivative of the equal time pair 
  amplitude.
\bibitem{sixth}In the region II, it is necessary to consider sixth-order 
  terms in the bulk free energy in order to lift the degeneracy present up to
  the fourth order.
\end{references}
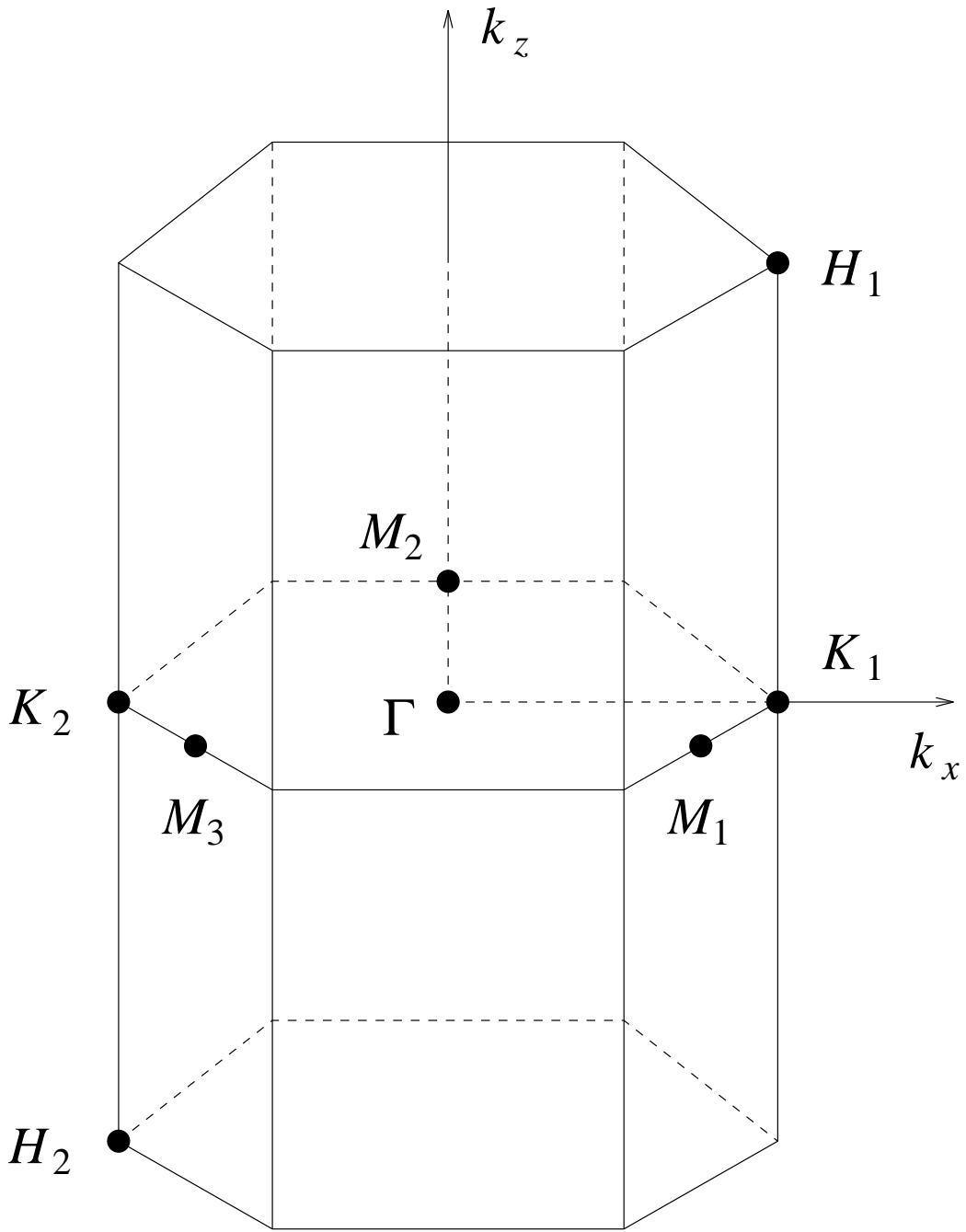
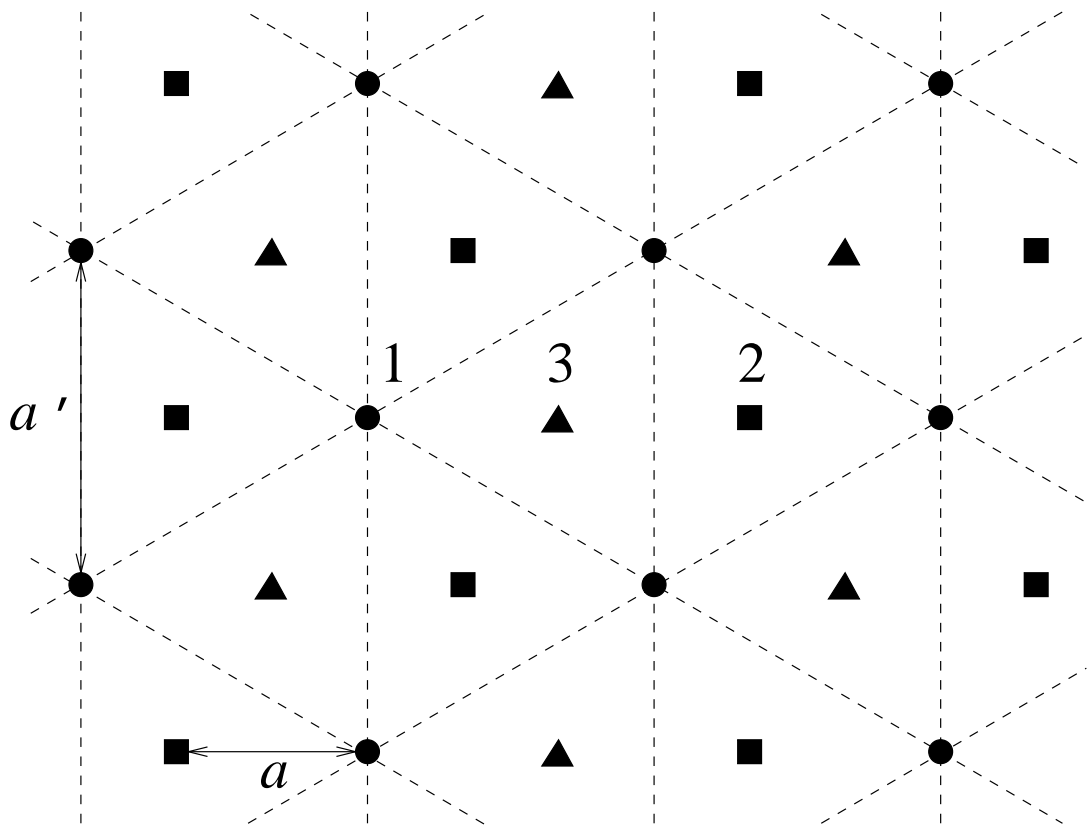
\end{document}